\documentclass{article}

\PassOptionsToPackage{numbers, sort&compress}{natbib}
 \usepackage[final]{tackling_climate_workshop_style}

\usepackage[utf8]{inputenc}    % allow utf-8 input
\usepackage[T1]{fontenc}       % use 8-bit T1 fonts

\usepackage{hyperref}

\usepackage{url}               % simple URL typesetting
\usepackage{booktabs}          % professional-quality tables
\usepackage{amsfonts}          % blackboard math symbols
\usepackage{nicefrac}          % compact symbols for 1/2, etc.
\usepackage{microtype}         % microtypography
\usepackage{graphicx}
\usepackage{amsmath}
\usepackage{float}
\usepackage{caption}
\usepackage{enumitem}
\usepackage{cleveref}

\title{\textit{EcoCast:} A Spatio-Temporal Model for Continual Biodiversity and Climate Risk Forecasting}

\author{%
  Hammed A.~Akande \\
  Department of Biology\\
  Concordia University\\
  Montreal, QC, Canada \\
  \texttt{akandehammedadedamola@gmail.com} \\
  \And
  Abdulrauf A.~Gidado \\
  Department of Computer Science and Mathematics\\
  Algoma University\\
  ON, Canada \\
  \texttt{abdulrauf.gidado@algomau.ca}
}

\begin{document}

\maketitle

\begin{abstract}
Increasing climate change and habitat loss are driving unprecedented shifts in species distributions. Conservation professionals urgently need timely, high-resolution predictions of biodiversity risks, especially in ecologically diverse regions like Africa. We propose \emph{EcoCast}, a spatio-temporal model designed for continual biodiversity and climate risk forecasting. Utilizing multisource satellite imagery, climate data, and citizen science occurrence records, \emph{EcoCast} predicts near-term (monthly to seasonal) shifts in species distributions through sequence-based transformers that model spatio-temporal environmental dependencies. The architecture is designed with support for continual learning to enable future operational deployment with new data streams. Our pilot study in Africa shows promising improvements in forecasting distributions of selected bird species compared to a Random Forest baseline, highlighting \emph{EcoCast's} potential to inform targeted conservation policies. By demonstrating an end-to-end pipeline from multi-modal data ingestion to operational forecasting, \emph{EcoCast} bridges the gap between cutting-edge machine learning and biodiversity management, ultimately guiding data-driven strategies for climate resilience and ecosystem conservation throughout Africa.
\end{abstract}

\section{Introduction}

 Climate change, in addition to widespread habitat degradation, is reshaping ecosystems across continental Africa, a threat intensified by increasing temperatures, changing precipitation patterns, and extreme weather events \cite{Sinatayehu2018}. These changing environments, largely driven by anthropogenic pressures, significantly affect biodiversity, particularly bird species that are highly sensitive to environmental change and serve as early warning indicators of ecosystem health \cite{moussy2022}. However, conservation professionals and policymakers often lack timely, fine-grained forecasts of changing species distribution or habitat quality over time. Current methods often involve static Species Distribution Models (SDMs) or climate risk assessments, which cannot keep pace with rapidly changing environmental conditions. In Africa, the lack of up-to-date biodiversity data has made it nearly impossible to track species declines or respond proactively \cite{Imarhiagbe2020}.

 We propose \emph{EcoCast}, a spatio-temporal model tailored to the dynamic nature of African ecosystems. \emph{EcoCast} is designed to ingest real-time and near-real-time data, from satellite imagery and climate data to citizen science observations, to generate frequently updated forecasts of shifting species distribution and habitat quality changes. By piloting \emph{EcoCast} in Africa, we illustrate how advanced machine learning can directly inform local conservation efforts, such as identifying emerging biodiversity hotspots, areas at risk of habitat loss, or assessing the ecological impact of infrastructure projects. In doing so, we aim to demonstrate a new paradigm where a single large-scale model can serve as a continuously improving ``environmental model'' for conservation decisions.

 \subsection{Key Contributions}

We develop a large-scale pre-trained model that integrates multimodal data (satellite images, species occurrences, climate variables) for biodiversity forecasting. To our knowledge, \emph{EcoCast} is among the first transformer-based models applied to ecological and climate risk data in Africa, learning generalizable representations that transfer across species and regions. The architecture supports continuous fine-tuning that can adapt to new data streams without catastrophic forgetting \cite{Yu2024}, enabling \emph{EcoCast} to maintain high predictive performance in a changing environment. Furthermore, \emph{EcoCast} is designed to provide a user-friendly interface for conservation professionals, allowing stakeholders to visualize model forecasts, explore scenarios, and contribute new observations.

\section{Related Work}

Over the years, machine learning techniques have been used to address key environmental challenges, including forest cover change detection \cite{hansen2013high}, crop-yield prediction from remote sensing \cite{you2017deep}, and SDMs \cite{elith2009species}. Recent advances in spatio-temporal deep learning provide a foundation for our approach. In climate and Earth observation, transformer models (e.g., \emph{EarthFormer}) and ConvLSTM capture space--time dependencies for tasks such as precipitation and ENSO forecasting \cite{gao2022earthformer,shi2015convolutional}. \emph{EcoCast} builds on these advances by adopting a similar Transformer-based architecture, but extends it to the multi-modal context of ecological forecasting with a persistent model that can be incrementally retrained with new data to stay current. The framework incorporates rehearsal-style updates combined with stability mechanisms to mitigate forgetting and support incremental retraining \cite{rebuffi2017icarl}. Recent work in ecological forecasting distinguishes between long-term climate projection-based SDMs and near-term operational forecasts that inform immediate management decisions \cite{dietze2018iterative,anderson2020forecast}. EcoCast extends this paradigm to biodiversity applications through transformer-based sequence modeling, bridging the gap between geophysical forecasting architectures and species distribution modeling.

%\section{Proposed Approach}

\section{Proposed Approach}\label{sec:proposed}

\emph{EcoCast's} workflow is structured into three phases (Figure~\ref{fig:ecocast-architecture} provides a conceptual overview). First, we create a pre-trained spatio-temporal model using historical multimodal data. Next, the framework is designed with a continuous learning pipeline to ingest new data and update the model incrementally. Finally, we envision deployment through practitioner-focused tools for visualization and feedback.

We use a monthly environmental time series covering the period from 2016 to 2023. Satellite inputs are monthly Sentinel-2 composites derived from 30\,m products and resampled onto a fixed continent-scale grid, and climate inputs are monthly ERA5 variables (Appendix~\ref{app:covariates}). Bird occurrence data were obtained from  Global Biodiversity Information Facility (GBIF) and aggregated by month to a 0.1$^{\circ}$ grid ($\approx$ 10--11\,km at the equator) across Africa. We pre-train our model on data from 2016 to 2021 and fine-tune it on 2022 before evaluating on 2023. By pre-training on this broad dataset, the model learns general spatio-temporal patterns (such as seasonal vegetation cycles or climate anomalies) relevant to the species' ecology.

\subsection{Operational Near-term Forecasting Framework}

We train a transformer-based architecture for spatio-temporal sequence modeling. Unlike traditional SDMs that require projected future climate scenarios (e.g., RCP4.5, RCP8.5) to forecast decades ahead, EcoCast implements an operational near-term forecasting paradigm designed for real-time conservation decision-making at monthly to seasonal timescales (see Appendix~\ref{app:sdm-comparison} for detailed comparison).

At each time step $t$, the model receives a temporal sequence of environmental covariates $\mathbf{x}_{t-L+1:t} = (x_{t-L+1}, \ldots, x_t)$, where each monthly observation $x_\tau$ combines satellite imagery features (Sentinel-2 bands, vegetation indices) and climate variables (ERA5 temperature, precipitation, etc.). The architecture predicts species occurrence $y_{t+1}$ for the \textit{next} month using observed historical data (Equation~\ref{eq:objective}):

\begin{equation}
\label{eq:objective}
\min_{\theta} \sum_{(x_{t-L+1:t}, y_{t+1})} \mathcal{L}(f_\theta(\mathbf{x}_{t-L+1:t}), y_{t+1}),
\end{equation}

\noindent where $\mathcal{L}$ is a class-imbalance-robust loss and $f_\theta$ denotes our transformer encoder with parameters $\theta$. We set the sequence length $L=12$ months to capture annual seasonality critical for avian phenology while maintaining computational efficiency.

We summarize resampled Sentinel-2 inputs within each 0.1$^{\circ}$ grid cell per month using band-wise statistics, and similarly aggregate monthly ERA5 climate variables at the same spatial resolution. At each grid cell and month $t$, we combine satellite and climate features into a unified environmental vector $\mathbf{x}_t \in \mathbb{R}^F$, where $F$ includes spectral bands, vegetation indices, and climate variables (see Appendix~\ref{app:temporal-alignment} for detailed illustration). To capture temporal dynamics and seasonal patterns, the model ingests sequences of $L$ consecutive monthly vectors $(\mathbf{x}_{t-L+1}, \ldots, \mathbf{x}_t) \in \mathbb{R}^{L \times F}$ and predicts species occurrence $y_{t+1}$ at the next time step.

Given that our species data are presence-only, for training, we generate pseudo-absences within the study area and the same temporal window. We thin occurrences spatially to the 0.1$^{\circ}$ grid to reduce clustering and sampling bias\cite{phillips2009sample, barbet-massin2012}. Additionally, we compute a sampling-effort covariate (based on monthly GBIF records) to help the model account for areas of high observer effort. The same pseudo-absence and thinning protocol is applied to the baseline model for fairness. In addition, we use stratified mini-batches that balance presence and pseudo-absence instances per species and a class-weighted approach to stabilize learning on rare positives.

For continual learning, our framework is designed to perform constrained updates when new data at \(t{+}1\) arrive. The architecture supports rehearsal-based updates with weight regularization, maintaining a fixed-size replay buffer of past examples and applying Elastic Weight Consolidation (EWC) to prevent catastrophic forgetting. The fine-tuning objective combines supervised loss on new data with a regularization term (Equation~\ref{eq:fine-tuning}):

\begin{equation}
\label{eq:fine-tuning}
\min_{\theta'}\ \sum_{k=1}^{K}\mathcal{L}\!\big(f_{\theta'}(x_{t+k}),y_{t+k}\big)\ +\ \lambda\,\Omega_{\text{EWC}}(\theta',\theta),
\end{equation}

where \(\Omega_{\text{EWC}}\) penalizes deviations from important parameters estimated on prior data. For the current evaluation, we employ rolling-origin validation with warm-start initialization, where each temporal window trains on all available historical data up to that point. This simulates operational deployment conditions while providing a rigorous baseline for future continual learning work.

To interpret predictions, we compute transformer attention rollout over space--time to highlight influential regions and months, and apply patch-wise permutation tests to verify that attributions reflect signals. For additional diagnostics, we use Grad-CAM-style attribution adapted to ViTs \cite{abnar2020quantifying,chefer2021transformer}. We flag cases where attributions may contradict ecological priors (for instance, focusing on clouds rather than vegetation change). We adopt spatio-temporal block cross-validation where folds are defined by spatial blocks and held-out years to prevent leakage and over-optimism\cite{valavi2019blockcv}. Thresholds for F1 are selected on the validation split per species; AUC is computed per species and then macro-averaged across species.

\section{Results}

A pilot study was conducted on five bird species (see Appendix~\ref{app:species} for the full list) across Africa. The monthly Sentinel-2 and ERA5 inputs were prepared as described in Section~\ref{sec:proposed}, and the GBIF records were aggregated to 0.1$^{\circ}$ cells by month. As a baseline, we trained species-specific Random Forest classifiers using static covariates plus occurrence records, with the \emph{same} pseudo-absence, thinning, and block CV protocol as the main \emph{EcoCast} model.

\subsection{EcoCast Implementation and Results}

As described in Section~\ref{sec:proposed}, EcoCast performs sequence-to-point forecasting: 12-month environmental sequences (satellite + climate features) predict next-month species occurrence. This operational framework was trained on 2016--2021 data, fine-tuned on 2022, and evaluated on held-out 2023 observations using rolling-origin validation to simulate real-world deployment conditions.

Table~\ref{tab:results} demonstrates EcoCast's performance gains over the Random Forest baseline. The transformer architecture's ability to model long-range temporal dependencies via self-attention, learn species-specific phenological patterns through positional encoding, and leverage cross-species ecological signals through joint multi-label training yields macro-averaged F1 and PR-AUC improvements of +34 and +43 percentage points, respectively. These results validate that explicit sequence modeling through transformer architectures is critical for near-term biodiversity forecasting, providing a foundation for the continual learning extensions and multi-task predictions planned for future iterations (see Appendix~\ref{app:architectural-advantages} for detailed architectural analysis).

\begin{table}[ht]
\centering
\caption{Comparative performance on 2023 holdout data for five bird species. \textbf{RF-ROE} (Random Forest with Rolling Origin Evaluation) trains separate models per species using flattened tabular features without explicit temporal modeling. \textbf{EcoCast} employs a transformer encoder over environmental sequences to predict species occurrence at month $t+1$.}
\label{tab:results}
\begin{tabular}{lcc}
\toprule
\textbf{Model} & \textbf{F1 macro} & \textbf{PR AUC macro} \\
\midrule
Baseline (Random Forest ROE) & 0.31 & 0.29 \\
\textbf{EcoCast (t+1 forecast)} & \textbf{0.65} & \textbf{0.72} \\
\bottomrule
\end{tabular}
\end{table}

\section{Pathway to Impact}

We plan to partner with groups like the Wildlife Conservation Society to validate \emph{EcoCast} outputs on the ground. The Federal Ministry of Environment could use \emph{EcoCast} to inform protected area management. This model will provide real-time forecasts highlighting areas at the highest biodiversity risk, enabling swift allocation of resources. Additionally, this could guide scenario analysis for land-use planning, evaluating the potential impacts of infrastructure projects (for instance, new roads) on bird habitats, and inform more sustainable development strategies. Once validated in Africa, the approach can extend to other continent-wide biodiversity hotspots. We would develop a robust API to enable African conservation NGOs and researchers to integrate local data, thereby further improving \emph{EcoCast's} coverage and accuracy. Protecting biodiversity hotspots can sustain ecotourism and preserve ecosystem services, supporting livelihoods. \emph{EcoCast} can encourage citizen-science participation, raising ecological awareness and empowering local stewardship.

\section{Conclusion}

We presented \emph{EcoCast}, a spatio-temporal model for continual forecasting of biodiversity and climate-related risks. \emph{EcoCast} demonstrates how advances in machine learning, from powerful Transformer architectures to continual learning techniques, can be utilized to address emerging challenges in conservation. Our case study in Africa showed that \emph{EcoCast} substantially outperforms traditional static models in predicting near-term species distribution shifts, and more importantly, provides actionable insights to conservation practitioners.

This work bridges the gap between cutting-edge AI and traditional ecology. Our proposed model is designed as an evolving model, continuously refreshed with new data, and well-suited to addressing climate change's non-stationary impact on ecosystems. By integrating data and delivering user-centric tools, we ensure that the model's intelligence is accessible to important stakeholders who need it most – conservation practitioners making important conservation decisions. Moving forward, we aim to expand \emph{EcoCast's} coverage and involve a broader community of experts to carefully evaluate its predictions. We will also explore scaling up the model (both in terms of architecture and training data) to capture continental-scale patterns and to include other taxa beyond birds. Additionally, policy engagement will be important. Demonstrating success stories where \emph{EcoCast}-informed interventions helped protect a species or habitat will build confidence among policymakers to formally incorporate such tools into conservation planning processes. \emph{EcoCast} shows a promising pathway for AI for Good in the environmental domain. By continuously learning from the planet's data and looping humans in decision-making, such systems can help us stay one step ahead in preserving biodiversity and fostering global climate resilience.

%\medskip

\small

\bibliographystyle{unsrtnat} %plainnat or unsrtnat
\bibliography{tackling_climate_workshop}

%\break
\newpage

\appendix

\section{Appendix: Conceptual Diagram, Selected Species, and Environmental Variables}

\subsection{Conceptual architecture of EcoCast}
\label{fig:ecocast-architecture}

\begin{figure}[!htbp]
  \centering
  \includegraphics[width=0.75\linewidth]{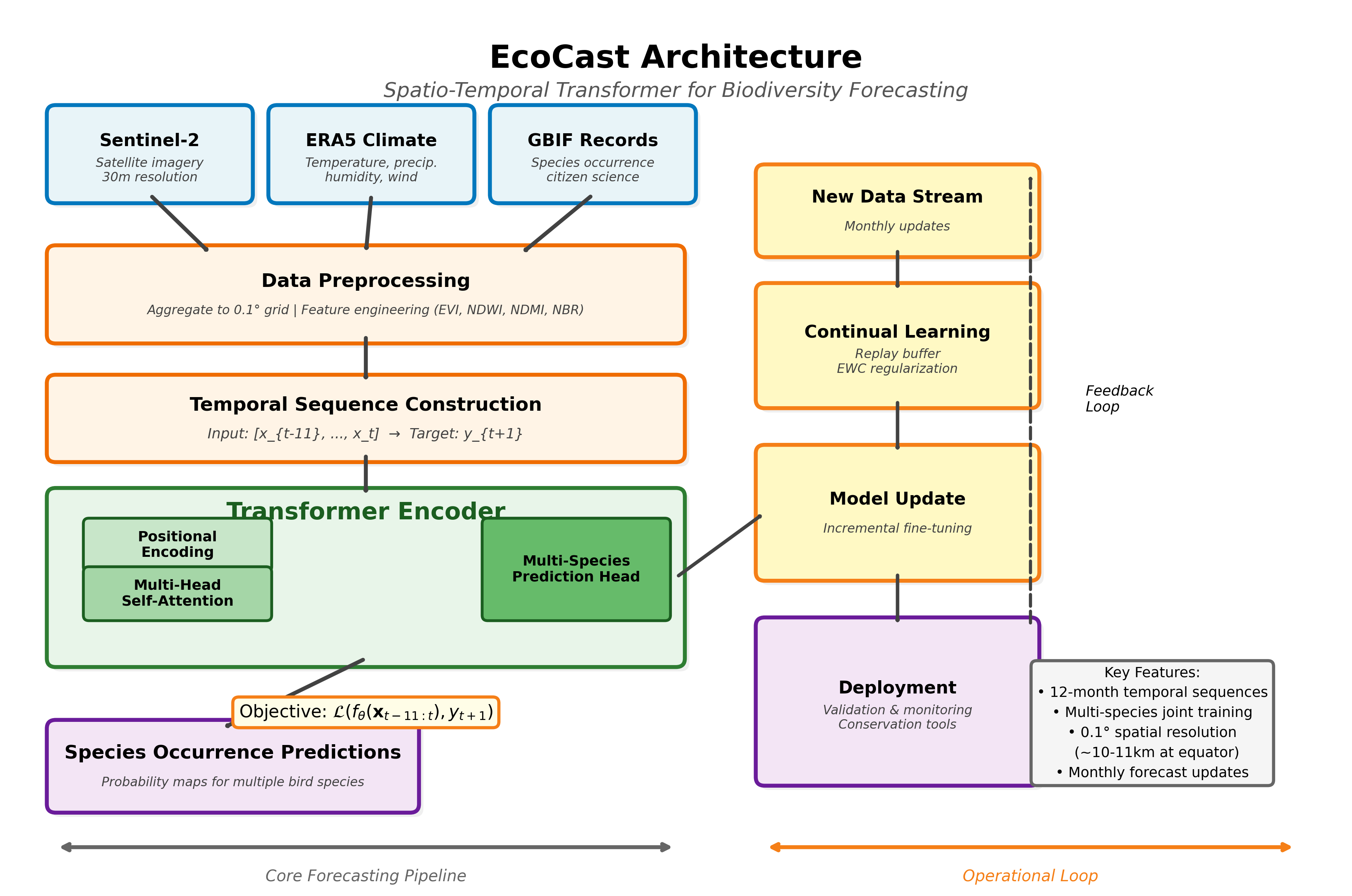}
  \caption{Conceptual architecture: \emph{EcoCast} ingests monthly Sentinel‑2 and ERA5 covariates aggregated to a 10\,km grid, stacks 12-month temporal sequences, and predicts multi‑species occurrence with a transformer encoder. The model performs $t + 1$ forecasting: given environmental features from months $t-11$ through $t$, it predicts species occurrence at month $t+1$ using self-attention mechanisms to capture temporal dependencies and seasonal patterns.}
  \phantomsection
  \label{fig:ecocast-architecture}
\end{figure}

\subsection{EcoCast Predicted Probability of Occurrence}
\label{fig:ecocast-sdm_1}

\begin{figure}[!htbp]
  \centering
  % Avoid spaces in filenames; rename the PNG like below
  \includegraphics[width=0.85\linewidth]{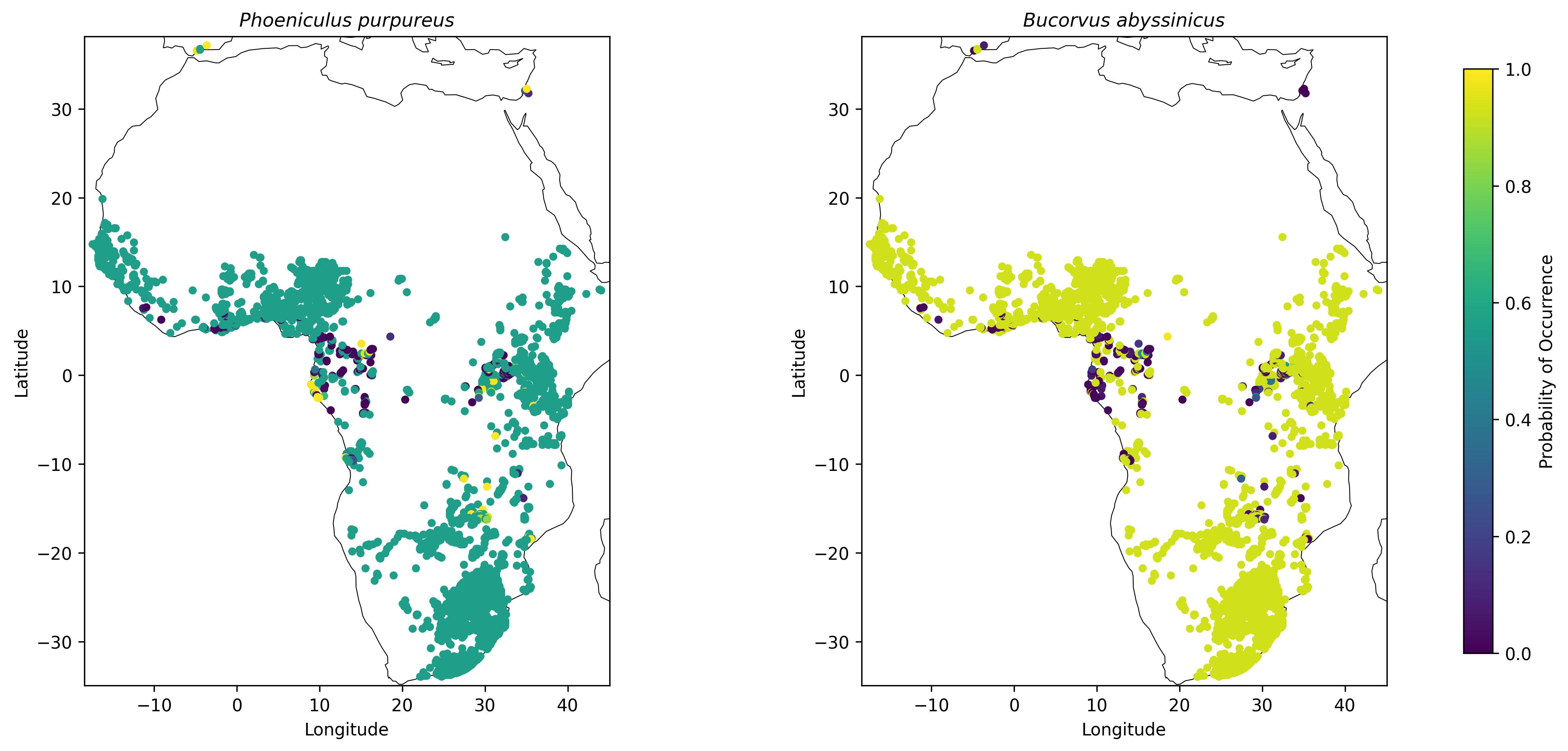}
  \caption{\textbf{EcoCast operational forecast (January 2024).}
  Predicted probability of occurrence across Africa for \emph{Phoeniculus purpureus}
  and \emph{Bucorvus abyssinicus}. The model uses 12-month environmental sequences
  (January--December 2023) to forecast species distributions in January 2024,
  demonstrating EcoCast's capability for operational near-term biodiversity forecasting.
  This represents a true out-of-sample forecast beyond the test period (2023),
  illustrating how conservation practitioners could use EcoCast to anticipate
  monthly shifts in species distributions.}
  \label{fig:ecocast-sdm_1}
\end{figure}

\newpage
\subsection{Species List}
\label{app:species}

The five focal species used in the pilot study are:
\begin{itemize}
    \item \emph{Psittacus erithacus} (African Grey Parrot)
    \item \emph{Pitta angolensis} (African Pitta)
    \item \emph{Tauraco persa} (Guinea Turaco)
    \item \emph{Bucorvus abyssinicus} (Northern Ground-Hornbill)
    \item \emph{Phoeniculus purpureus} (Green Wood-Hoopoe)
\end{itemize}

\subsection{Environmental Covariates}
\label{app:covariates}

For the Sentinel-2 composites, we derived band-wise statistics and vegetation indices: EVI, NDWI, NDMI, and NBR.

ERA5 climate variables included: temperature, relative humidity, total precipitation, wind speed, and surface pressure.

\subsection{Detailed Comparison with Traditional SDMs}
\label{app:sdm-comparison}

This approach fundamentally differs from climate projection-based SDMs in three ways:
\begin{enumerate}[leftmargin=*,itemsep=0pt]
    \item \textbf{No future climate projections required}: EcoCast learns temporal dependencies and lagged ecological responses from observed environmental sequences, enabling predictions \textit{without} requiring GCM outputs or emission scenario assumptions.

    \item \textbf{Sequence modeling via self-attention}: Unlike Random Forest baselines that treat each (location, month) observation independently, the transformer architecture explicitly models inter-month dependencies through multi-head self-attention. This captures phenological patterns (e.g., ``vegetation green-up in month $t-3$ predicts migratory arrival at $t$'') and lagged climate effects automatically, without manual feature engineering.

    \item \textbf{Near-real-time operational capability}: ERA5 climate reanalysis becomes available within 5 days of observation (preliminary ERA5T) with final quality-assured data released 2--3 months later \cite{copernicus2025era5}. This enables monthly forecast updates aligned with conservation planning cycles, rather than static decadal projections.
\end{enumerate}

Our architecture supports longer windows ($L \leq 24$) when deeper historical context is available.

\subsection{Architectural Advantages}
\label{app:architectural-advantages}

The performance gap between EcoCast and RF-ROE highlights the limitations of treating temporal biodiversity data as independent tabular observations. Random forests, while powerful for static prediction tasks, lack mechanisms to capture:
\begin{itemize}[leftmargin=*,itemsep=0pt]
    \item \textbf{Temporal autocorrelation}: Bird presence in month $t$ is highly predictive of month $t+1$ for resident species, yet RF models each month independently.
    \item \textbf{Lagged environmental responses}: Species may respond to climate anomalies with 2--4 month delays (e.g., food availability following rainfall), which requires sequence modeling to capture effectively.
    \item \textbf{Seasonal periodicity}: Migratory species exhibit annual cycles that transformers encode through positional embeddings, whereas RF requires manual construction of seasonal features.
\end{itemize}

\subsection{Glossary of Acronyms}
\label{app:glossary}

\begin{itemize}
    \item \textbf{ERA5}: Fifth-generation of the European Centre for Medium-Range Weather Forecasts (ECMWF) reanalysis for global climate and weather.
    \item \textbf{SDM}: Species Distribution Model
    \item \textbf{GBIF}: Global Biodiversity Information Facility, an international database of species occurrence records.
    \item \textbf{ViT}: Vision Transformer, a Transformer-based architecture for image classification and analysis.
    \item \textbf{Grad-CAM}: Gradient-weighted Class Activation Mapping.
    \item \textbf{EVI}: Enhanced Vegetation Index, a vegetation index that corrects for atmospheric and soil background effects.
    \item \textbf{NDWI}: Normalized Difference Water Index, highlights water content in vegetation and soil.
    \item \textbf{NDMI}: Normalized Difference Moisture Index, sensitive to vegetation canopy water content.
    \item \textbf{NBR}: Normalized Burn Ratio, used to detect burned areas and vegetation disturbance.
    \item \textbf{ConvLSTM}: Convolutional Long Short-Term Memory network, a recurrent neural network that combines convolutional layers with LSTMs to capture spatio-temporal patterns.
    \item \textbf{EarthFormer}: A spatio-temporal Transformer model developed for Earth system forecasting
    \item \textbf{Block CV}: Block cross-validation, a spatial or spatio-temporal cross-validation scheme.
    \item \textbf{ROE}: Rolling Origin Evaluation. In this work, we pair ROE with Random Forest (RF-ROE) as a baseline species distribution model.

\end{itemize}

\subsection{Temporal Alignment of Satellite and Climate Data}
\label{app:temporal-alignment}

This section details the temporal alignment of satellite imagery and climate variables within EcoCast's sequence modelling framework. \textbf{Both data sources undergo identical temporal processing}: satellite and climate features are concatenated at each monthly time step to form unified environmental vectors, which are then arranged into temporal sequences for input to the transformer model.

\subsubsection{Unified Temporal Processing}

\begin{figure}[!htbp]
  \centering
  \includegraphics[width=0.95\linewidth]{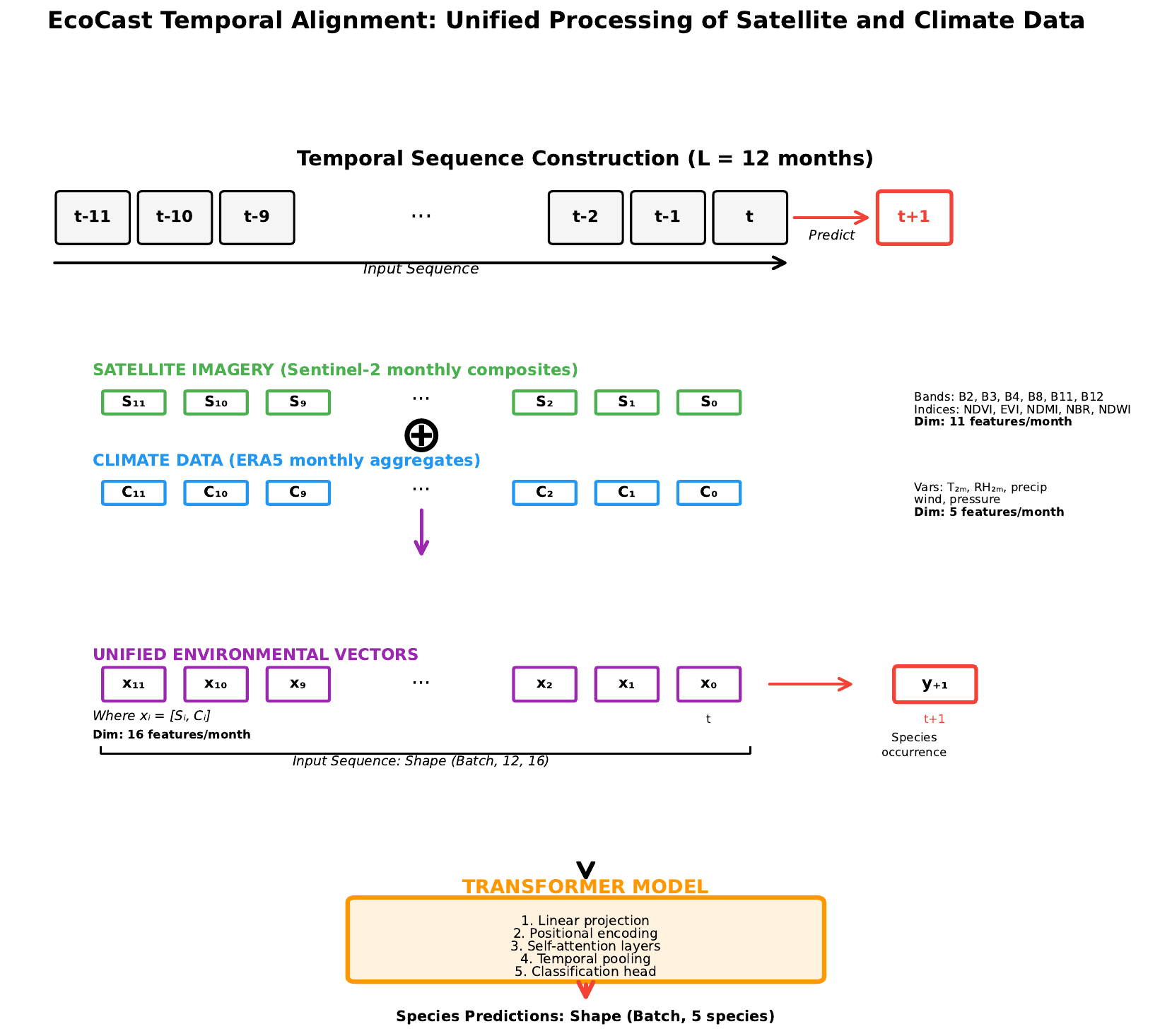}
  \caption{\textbf{Temporal alignment diagram for EcoCast.} Both satellite imagery (Sentinel-2) and climate data (ERA5) are processed as monthly sequences with lag $L=12$ months. At each time step, satellite and climate features are concatenated ($\oplus$) into unified environmental vectors $\mathbf{x}_t$ before sequence construction. The transformer model attends over this 12-month input sequence to predict species occurrence at month $t+1$. This design ensures both modalities contribute equally to temporal dependency learning through self-attention mechanisms.}
  \label{fig:temporal-alignment}
\end{figure}

\emph{EcoCast} constructs environmental sequences by concatenating satellite and climate features \textit{at each monthly time step} before sequence construction. At each grid cell and month $t$, we combine:

\begin{equation}
\mathbf{x}_t = [\mathbf{s}_t, \mathbf{c}_t] \in \mathbb{R}^{16}
\end{equation}

\noindent where $\mathbf{s}_t \in \mathbb{R}^{11}$ contains Sentinel-2 bands (B2, B3, B4, B8, B11, B12) and vegetation indices (NDVI, EVI, NDMI, NBR, NDWI) for month $t$, and $\mathbf{c}_t \in \mathbb{R}^{5}$ contains ERA5 climate variables (2m temperature, relative humidity, precipitation, wind speed, surface pressure) for the same month. The model then ingests a sequence of these unified vectors:

\begin{equation}
\mathbf{X} = [\mathbf{x}_{t-L+1}, \ldots, \mathbf{x}_t] \in \mathbb{R}^{L \times 16}
\end{equation}

\noindent with $L=12$ months, to predict species occurrence $y_{t+1}$ at the subsequent time step. Figure~\ref{fig:temporal-alignment} illustrates this construction process.

\subsubsection{Key Design Principles}

Three critical aspects ensure consistent temporal treatment:

\begin{enumerate}[leftmargin=*,itemsep=2pt]
    \item \textbf{Matched temporal resolution}: Both satellite composites and climate aggregates are computed at monthly granularity. Sentinel-2 observations within month $t$ are mosaicked and cloud-masked to produce representative monthly statistics, while ERA5 hourly data are averaged to monthly means.

    \item \textbf{Aligned spatial resolution}: All features are resampled to a 0.1$^\circ$ grid ($\approx$10--11\,km at the equator) before concatenation, ensuring spatial correspondence between satellite pixels and climate variables.

    \item \textbf{Uniform lag structure}: The 12-month lookback window applies to the \textit{combined} feature vector $\mathbf{x}_t$, not to satellite and climate data separately. The transformer's self-attention mechanism operates over $[\mathbf{x}_{t-11}, \ldots, \mathbf{x}_t]$ as a unified sequence.
\end{enumerate}

\subsubsection{Architectural Implementation}

EcoCast's architecture is designed to learn joint spatio-temporal patterns across modalities. Formally, sequences are constructed \textit{after} feature concatenation at each time step:

\begin{equation}
[\mathbf{x}_{t-11}, \ldots, \mathbf{x}_t] = [[\mathbf{s}_{t-11}, \mathbf{c}_{t-11}], \ldots, [\mathbf{s}_t, \mathbf{c}_t]] \longrightarrow y_{t+1}
\end{equation}

\noindent This design enables the transformer to learn cross-modal temporal dependencies through self-attention. For example, the model can learn that vegetation greenness (satellite) in month $t-3$ combined with precipitation (climate) in months $t-4$ to $t-2$ jointly predict species arrival at month $t+1$, without requiring manual specification of these lagged interactions.

\subsubsection{Notation Summary}

For clarity, we provide consolidated notation:

\begin{itemize}[leftmargin=*,itemsep=1pt]
    \item $t$: Current time step (month)
    \item $L = 12$: Sequence length (lookback window)
    \item $\mathbf{s}_t \in \mathbb{R}^{11}$: Satellite features at month $t$
    \item $\mathbf{c}_t \in \mathbb{R}^{5}$: Climate features at month $t$
    \item $\mathbf{x}_t = [\mathbf{s}_t, \mathbf{c}_t] \in \mathbb{R}^{16}$: Unified environmental vector
    \item $\mathbf{X} = [\mathbf{x}_{t-11}, \ldots, \mathbf{x}_t] \in \mathbb{R}^{12 \times 16}$: Input sequence
    \item $y_{t+1} \in \{0,1\}^5$: Multi-species occurrence at next month (5 bird species)
\end{itemize}

This uniform treatment ensures that both satellite imagery and climate data contribute equally to the model's understanding of spatio-temporal environmental dynamics, enabling robust near-term biodiversity forecasting without requiring manual feature engineering or separate temporal modeling pipelines.

\end{document}